\documentclass[11pt,preprint]{aastex}
\usepackage{apjfonts}

\def\gax{\mathrel{\raise.3ex\hbox{$>$}\mkern-14mu\lower0.6ex\hbox{$\sim$}}}
\def\lax{\mathrel{\raise.3ex\hbox{$<$}\mkern-14mu\lower0.6ex\hbox{$\sim$}}}
\def\gtorder{\mathrel{\raise.3ex\hbox{$>$}\mkern-14mu
             \lower0.6ex\hbox{$\sim$}}}
\def\ltorder{\mathrel{\raise.3ex\hbox{$<$}\mkern-14mu
             \lower0.6ex\hbox{$\sim$}}}

\begin{document}

\title{Stellar Binary Companions to Supernova Progenitors}

\author{Christopher S.~Kochanek\altaffilmark{1,2}}
 
\altaffiltext{1}{Department of Astronomy, The Ohio State University, 140 West 18th Avenue, Columbus OH 43210}
\altaffiltext{2}{Center for Cosmology and AstroParticle Physics, The Ohio State University, 191 W. Woodruff Avenue, Columbus OH 43210}

\begin{abstract}
For typical models of binary statistics, 50--70\% of core-collapse supernova 
(ccSN) progenitors are members of a stellar binary at the time of the explosion.  
Independent of any consequences of mass transfer, this has observational consequences 
that can be used to study the binary properties of massive stars.  In
particular, the secondary companion to the progenitor of a Type~Ib/c
SN is frequently ($\sim 50\%$) the more optically luminous star since the high effective
temperatures of the stripped progenitors make it relatively easy for
a lower luminosity, cooler secondary to emit more optical light.  Secondaries
to the lower mass progenitors of Type~II SN will frequently produce
excess blue emission relative to the spectral energy distribution of
the red primary.  Available data constrain the models weakly.  Any detected
secondaries also provide an independent lower bound on the progenitor mass
and, for historical SN, show that it was not a Type~Ia event.  Bright ccSN 
secondaries have an unambiguous, post-explosion observational signature -- 
strong, blue-shifted, relatively broad absorption lines created by 
the developing supernova remnant.  These can be used to locate historical 
SN with bright secondaries, confirm that a source is a secondary, and, potentially, 
measure abundances of ccSN ejecta.  Luminous, hot secondaries will reionize
the SNR on time scales of 100-1000~yrs that are faster than reionization by
the reverse shock, creating peculiar HII regions due to the high
metallicity and velocities of the ejecta.
\end{abstract}


\section{Introduction}
\label{sec:introduction}

The discovery of the blue progenitor to SN~1987A \citep{Gilmozzi1987} triggered interest
in the binarity of core-collapse supernova (ccSN) progenitors, since mass transfer 
provided a simple explanation for how a relatively low mass $\sim 15-20M_\odot$ star could 
explode as a blue supergiant  \citep{Podsiadlowski1989,Deloore1992}.  Later population
studies showed that $\sim 25\%$ of masssive stars may undergo sufficient mass
transfer to transform a primary that would explode as a red supergiant in a 
Type~II SN into a stripped Helium (or beyond) star that would explode as a 
Type~Ib/c SN  \citep[e.g.,][]{Podsiadlowski1992,Eldridge2008}.  If the secondary
gains enough mass as part of this process, its evolution accelerates and it can 
explode as a Type~II SN while still a blue supergiant.  

Many aspects of these possibilities were confirmed by the Type~IIb SN~1993J in 
M81 ($3.6$~Mpc).  A binary companion was first
suggested based on excess blue emission in the spectral energy distribution 
(SED) of the progenitor \citep{Aldering1994}, and later confirmed in post-explosion imaging and
spectroscopic observations \citep{Maund2004,Maund2009}.   Binary evolution
models produce SN~1993J starting with two similar mass stars \citep{Podsiadlowski1993,Maund2004,Stancliffe2009}.
Mass transfer removes almost all the Hydrogen envelope of the primary prior
to the explosion, leading to the Type~IIb transition from a Type~II to a Type~Ib spectral
type as the SN evolves.  

While direct searches for SN progenitors are now common, direct constraints on
binary companions are relatively rare.  \cite{Smartt2009b} reviews the 
status of progenitor searches \citep[e.g.,][]{2007ApJ...661.1013L,2007ApJ...656..372G,
Smartt2009}.  For Type~IIP SN there are 20 events with adequate pre-SN data,
with 6 detections and 12 strong upper bounds.  These progenitors are red
supergiants with $7M_\odot \ltorder M_p \ltorder 17M_\odot$, which is 
surprising given that locally red supergiants are found up to masses of
$25M_\odot$ \citep{Levesque2005}.  No progenitor of a Type~Ib/c SN has been
identified out of 10 cases with adequate pre-explosion data, which may
be further evidence that mass transfer is playing an important role in creating
these SN \citep{Smartt2009b}.  

Because the data on SN progenitors rarely have broad spectral coverage and
there is little adequate post-outburst data, there are only a handful
of constraints on companion stars other than those for SN~1993J.  Two
additional Type~IIb SN, SN~2008ax  \citep{Crockett2008} and SN~2001ig
\citep{Ryder2006}, show evidence for a secondary.  For SN~2008ax, there
is a blue excess to the SED of the progenitor, while for SN~2001ig there
is a candidate for the surviving, blue secondary in post explosion images.  
The existence of a companion is limited to $M_s \ltorder 20M_\odot$ for 
the Type~Ic SN~2002ap \citep{Crockett2007}, which also has the tightest 
limits on the progenitor of any Type~Ib/c SN, to $L \ltorder 2L_\odot$ for 
SN~1987A \citep{Graves2005}, and to roughly $M_V\gtorder 2$~mag for the 
Type~IIb SN Cassiopeia A \citep{Fesen2006,Krause2008}.  
Finally, the Type~IIP SN~2008bk \citep{Mattila2008} and SN~2005cs \citep{Maund2005,Li2006}
have upper bounds on excesses to their SEDs at V-band of approximately 
1~mag.  That the constraints are few is not surprising, given that it will
generally be more challenging to identify the secondaries.  We should note 
that there is also indirect evidence for the presence of a binary companion
from the structure of some radio supernova light curves \citep[see][]{Vandyk2004}.

\begin{figure}[t]
\plotone{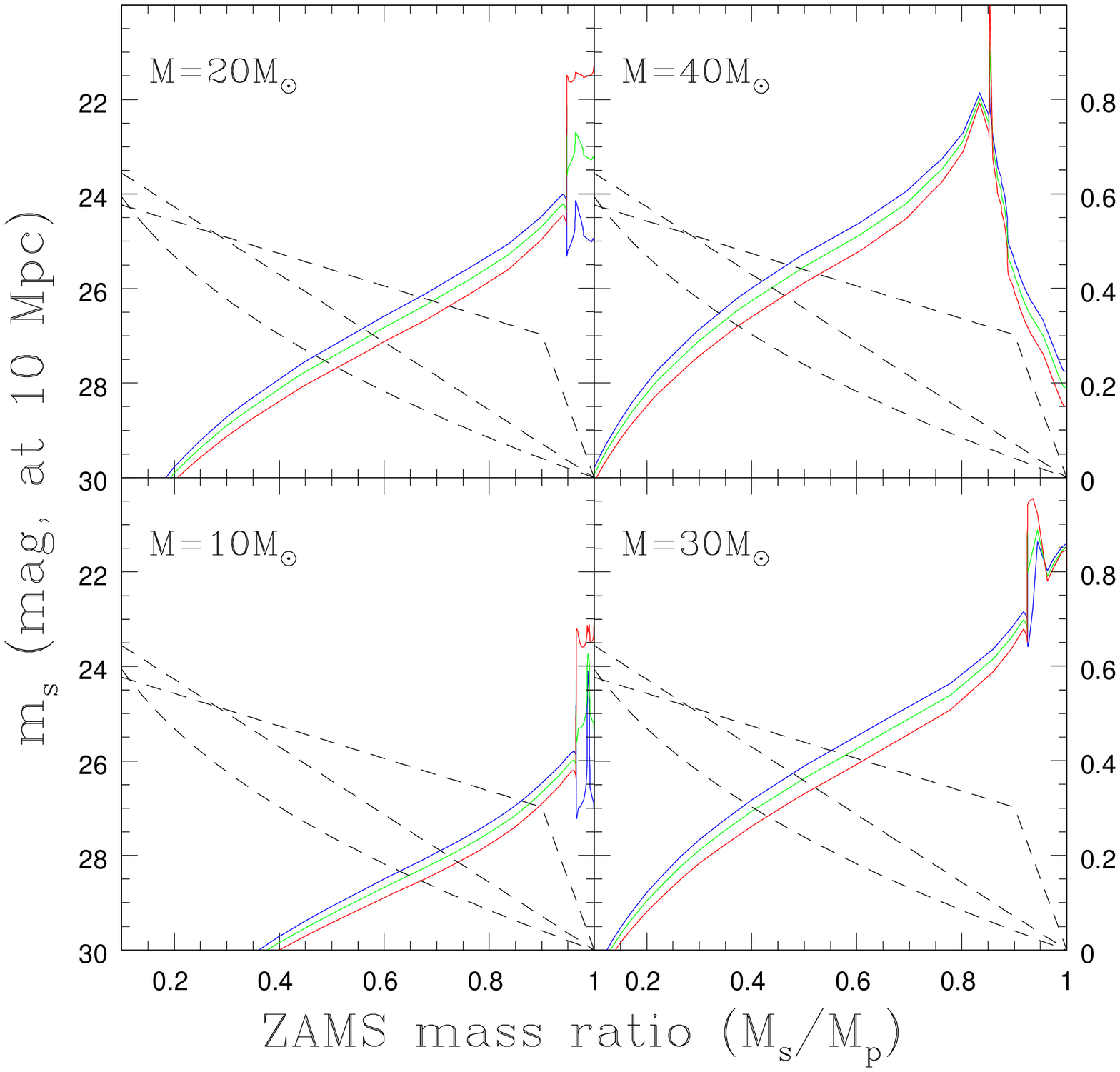}
\caption{
 Secondary magnitudes at the time of primary core collapse as a function of 
 the initial mass ratio $M_s/M_p$.  The magnitudes of the primary are the
 values for $M_s/M_p=1$.  
 The B (blue), V (green) and I (red)
 magnitudes are scaled to a distance of 10~Mpc.  The dashed lines show
 the integral distributions of the secondaries in the mass ratio 
 including the dilution by collapsing secondaries for binary fraction 
 $F=1$  (right scale). From top to bottom at high mass fractions they
 are the twins, uniform and low mass distributions.  To rescale these
 to a different binary fraction, multiply by the ratio $f_p(F)/f_p(F=1)$
 from Table~\ref{tab:fractions}.  For the \cite{Eldridge2008} evolution
 models, the $30 M_\odot$ case would resemble the $40 M_\odot$ case.
 }
\label{fig:mass}
\end{figure}

\begin{figure}[t]
\plotone{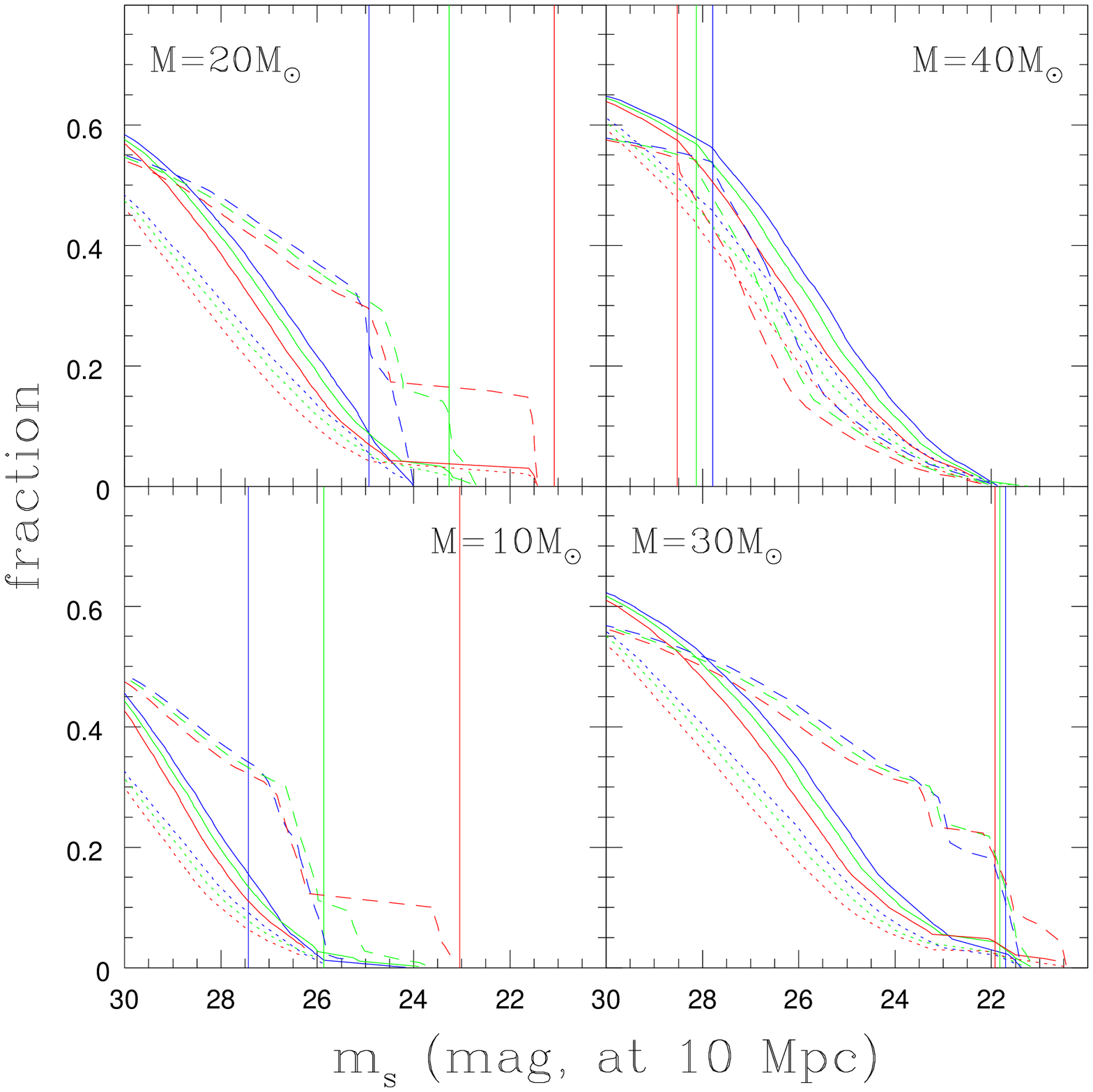}
\caption{
 Integral distributions of companion magnitudes in the B (blue), V (green) and
 I (red) at a distance of 10~Mpc.  The low mass (dotted), uniform (solid)
 and twins (dashed) distributions are shown for a binary fraction $F=1$ 
 and the diluting effects of collapsing secondaries are included.  To rescale 
 these to a different binary fraction, multiply by the ratio $f_p(F)/f_p(F=1)$
 from Table~\ref{tab:fractions}.  The vertical lines mark the magnitudes of
 the primary at the time of collapse.  Note how the intrinsically most
 luminous $M_p=40M_\odot$ progenitor is optically fainter than the three
 lower mass progenitors. For the \cite{Eldridge2008} evolution
 models, the $30 M_\odot$ case would resemble the $40 M_\odot$ case.
 }
\label{fig:frac1}
\end{figure}

Five years ago, limits and detections of SN progenitors would have been
equally limited, but the passage of time renders all studies of SN 
progenitors and their secondaries easier, since both the number of nearby
SN and the amount of useful archival data continuously increase.  This
suggests that an examination of our expectations for the properties of
binary companions to ccSN progenitors and of any strategies for
detection are worthwhile.  In this 
paper we first  estimate the fraction of ccSN that have a stellar binary 
companion at the time of the SN in \S\ref{sec:model} and then examine
the optical properties of these companions in \S\ref{sec:results}. 
In \S\ref{sec:observations} we examine whether these models are constrained
by any of the existing data on SN progenitors and their companions.
In \S\ref{sec:verify} we point out that there is a simple technique to discover
and unambiguously confirm the brighter secondary companions to ccSN
even when the position of the ccSN is poorly constrained.
Finally, in \S\ref{sec:discuss} we summarize the results and discuss
the observational prospects.  We examine these questions in the 
``wide binary'' limit, where there have been no interactions between the two stars.
In most cases, adding the interactions will enhance the visibility
of the secondary, as mass transfer frequently leads to a hotter primary 
radiating less in the optical and a more massive and luminous
secondary.  We will consider these effects in Kochanek et al. 
(2009, in preparation).  In an Appendix, we discuss other possibilities
for locating historical supernovae sufficiently accurately to search
for surviving secondaries.  

\section{The Fraction of ccSN in Stellar Binaries}
\label{sec:model}

We assume that the distribution of primary masses $M_p$ is Salpeter,
$dN/dM_p \propto M_p^{-x}$ with $x=2.35$, and that fraction $F$ of the
primaries have binary companions distributed in mass $M_s$ as $f(q)$
with $q_{min} \leq q=M_s/M_p \leq q_{max}$ where $\int dq f(q) \equiv 1$.
With these definitions, the joint distribution of primary
and secondary stars in mass is
\begin{equation}
   { d N \over dM_p dM_s } = F M_p^{-x-1} f(q),
\end{equation}
and the distribution of secondary masses is
\begin{equation}
   { d N \over dM_s } = F f_q M_s^{-x}
\end{equation}
where 
\begin{equation} 
      f_q =  \int_{q_{min}}^{q_{max}} q^{x-1} f(q) dq.
\end{equation}
Note that the distribution of secondaries in mass is also Salpeter
in form, and that at all masses there are $F f_q$ secondaries for
every primary.  When we observe a ccSN, it can be the collapse of 
a single star that was never in a binary, the primary of a binary
or the secondary of a binary, and the relative probabilities of
these three cases are   
\begin{equation}
   f_{single} = { 1-F \over 1 + F f_q } \quad\hbox{and}\quad  
   f_p = { F \over 1 + F f_q } \quad\hbox{and}\quad  
   f_s = { F f_q \over 1 + F f_q }.
\label{eqn:fractions}
\end{equation} 
As long as the primary mass function is a power-law and the secondary
mass function depends only on $q$, these fractions are mass-independent.
The fraction of ccSN in stellar binaries is simply $f_p$.

We consider three models for the distribution of secondaries.  The first
two are based on \cite{Kobulnicky2007} and are limited to $0.02 < q < 1$.  
The ``low mass'' model, $f(q) \propto q^{-0.6}$, is biased to low mass 
binary companions, while the ``uniform'' model with $f(q)$ constant 
gives equal weight to all masses.   
The third, ``twins'' model is based on \cite{Stanek2007} and 
puts half of the companions in a
twin distribution with $0.9 < q < 1$  and the other half with
$0.02 < q < 0.9$, using uniform distributions for each mass ratio interval.
The fractions of stars that are secondaries at a fixed mass
if $F=1$ are $f_q=0.288$, $f_q=0.434$, and $f_q=0.644$ for the low
mass, uniform and twins models, respectively. 
Table~\ref{tab:fractions} summarizes the fractions (Eqn.~\ref{eqn:fractions})
of ccSN that will be isolated stars, binary primaries and binary
secondaries for each of these models and binary fractions of
$F=1$, $3/4$, $1/2$ and $1/4$.
Biasing the binary distribution to lower masses increases $f_p$ 
because there are fewer, diluting massive secondaries. Reducing the
binary fraction $F$, particularly for high binary fractions, has a weaker
effect on $f_p$, because the
change is partly compensated by the reduced numbers of
exploding secondaries. If the binary fraction $F\gtorder 0.7$, 
as favored by  \cite{Kobulnicky2007}, then 50--80\% of ccSN 
are part of a stellar binary at the time of the explosion.

\begin{deluxetable}{rcccc}
\tablewidth{0pt}
\tablecaption{Binary Status at Time of ccSN}
\tablehead{
   \colhead{Model} &\colhead{F}  &\colhead{$f_{single}$}
  &\colhead{$f_p$} &\colhead{$f_s$}
  }
\startdata
  low mass &1.00 &0.00 &0.78 &0.22 \\
           &0.75 &0.21 &0.62 &0.18 \\
           &0.50 &0.44 &0.44 &0.13 \\
           &0.25 &0.70 &0.23 &0.07 \\
  uniform  &1.00 &0.00 &0.70 &0.30 \\
           &0.75 &0.19 &0.57 &0.25 \\
           &0.50 &0.41 &0.41 &0.18 \\
           &0.25 &0.67 &0.23 &0.10 \\
  twins    &1.00 &0.00 &0.60 &0.40 \\
           &0.75 &0.17 &0.50 &0.33 \\
           &0.50 &0.38 &0.38 &0.25 \\
           &0.25 &0.64 &0.21 &0.14 \\
\enddata
\tablecomments{The Model column defines the $f(q)$ distribution of the
  secondary stars (see text), $F$ is the fraction of primaries that are in
  binaries and $f_{single}$, $f_p$ and $f_s$ are the fraction of
  ccSN that are isolated stars, binary primaries and binary secondaries,
  respectively. }
\label{tab:fractions}
\end{deluxetable}

\begin{deluxetable}{rccccccccc}
\tablewidth{0pt}
\tablecaption{Observability of Binary Companions} 
\tablehead{
   \colhead{Model} 
       &\colhead{$M_p$}  
       &\multicolumn{2}{c}{B band}
       &\multicolumn{2}{c}{I band}
       &\colhead{$\hbox{I}_s\leq26$}
       &\colhead{$\hbox{I}_s\leq26$}
       &\colhead{$\hbox{I}_s\leq24$}
       &\colhead{$\hbox{I}_s\leq24$} \\
       &\colhead{$M_\odot$}  
       &\colhead{$>10\%$}
       &\colhead{$>100\%$}
       &\colhead{$>10\%$}
       &\colhead{$>100\%$}
       &\colhead{(10~Mpc)}
       &\colhead{(3~Mpc)}
       &\colhead{(10~Mpc)}
       &\colhead{(3~Mpc)}
  }
\startdata
  low mass &10 &0.32    &0.09    &0.02 &0.00 &0.02   &0.15 &0.02  &0.03\\
           &20 &0.26    &0.05    &0.02 &0.00 &0.09   &0.33 &0.02  &0.14\\
           &30 &0.09    &0.02    &0.07 &0.02 &0.17   &0.42 &0.06  &0.23\\
           &40 &0.63    &0.46    &0.65 &0.47 &0.21   &0.48 &0.06  &0.28\\
  uniform  &10 &0.45    &0.15    &0.02 &0.00 &0.02   &0.25 &0.02  &0.05\\
           &20 &0.38    &0.09    &0.04 &0.00 &0.16   &0.45 &0.04  &0.22\\
           &30 &0.14    &0.03    &0.12 &0.04 &0.28   &0.54 &0.09  &0.35\\
           &40 &0.65    &0.56    &0.66 &0.57 &0.29   &0.58 &0.09  &0.37\\
  twins    &10 &0.48    &0.34    &0.10 &0.00 &0.10   &0.38 &0.10  &0.22\\
           &20 &0.45    &0.24    &0.16 &0.00 &0.34   &0.48 &0.16  &0.37\\
           &30 &0.34    &0.12    &0.33 &0.19 &0.40   &0.52 &0.32  &0.44\\
           &40 &0.58    &0.54    &0.59 &0.54 &0.14   &0.55 &0.05  &0.23\\
\enddata
\tablecomments{The model column defines the $f(q)$ distribution of the
  secondary stars (see text), $M_p$ is the primary mass, the B band
  and I band columns give the fraction of secondaries exceeding 10\%
  and 100\% of the primary's flux in that band.
  The remaining columns show the fraction of SN with
  secondaries brighter than either $\hbox{I}=26$ or $24$~mag
  at either 10~Mpc or 3~Mpc.
  These are calculated for $F=1$ but can be corrected to lower 
  values of $F$ using Table~\ref{tab:fractions}.
   }
\label{tab:observe}
\end{deluxetable}

\section{Observability}
\label{sec:results}

That companions are common does not imply that they are observable, so
we next make a simple model of their visibility.  We took the \cite{Marigo2008}
solar metallicity isochrones, called the most massive 
star along the isochrone the SN progenitor, and then selected secondaries from 
the less massive stars along the same isochrone.  Massive stars reach the main
sequence rapidly compared to their post-main sequence life times \citep[e.g.,][]{Zinnecker2007}, 
so we can ignore small differences in the pre-main sequence evolution
of the binary stars and simply start from the zero age main sequence (ZAMS).
We used the estimated HST/ACS HST/ACS B (F435W), V (F555W) and I (F814W) 
magnitudes.  We  use the same binary distributions as in \S\ref{sec:model}
and assume $F=1$.  The resulting distributions can be scaled to lower binary
fractions using Table~\ref{tab:fractions}.  

We also examined the Geneva \citep{Lejeune2001} and Cambridge STARS \citep{Eldridge2008} model
sequences, but were able to use them only to spot check the results.
\cite{Lejeune2001} provide end-phase optical magnitudes only for 
$M\ltorder 25M_\odot$, while \cite{Eldridge2008} provide histories
for discrete masses that are too coarsely sampled to turn into 
isochrones.  Where we can compare results between \cite{Marigo2008} and 
\cite{Lejeune2001}, they are similar.  The final phases of the
\cite{Marigo2008} and   \cite{Eldridge2008} models are generally
similar over the mass range $10 M_\odot \ltorder M \ltorder 50M_\odot$,
but with significant differences in the onset of extreme mass loss.
In the \cite{Eldridge2008} models, the transition occurs abruptly at
$M\simeq 28M_\odot$, while in the \cite{Marigo2008} models there
is a steady shift over the range from $20 M_\odot < M < 40M_\odot$.
Thus, the $M_p=30M_\odot$ case we construct from \cite{Marigo2008} resembles
the $M_p=20 M_\odot$ case, while in the \cite{Eldridge2008} models
it would resemble the $M_p=40 M_\odot$ case.

Figure~\ref{fig:mass} shows the distributions of secondaries relative to
primaries in mass and magnitude, along with the integral distribution of
secondaries in mass for the $F=1$ models, including the dilution by the
ccSN of secondaries.  To scale these fractions to a different $F$, simply
multiply by the ratio $f_p(F)/f_p(F=1)$ from Table~\ref{tab:fractions}.  We show the
distributions for primary masses of $M_p\simeq 10$, $20$, $30$ and 
$40M_\odot$ and scale the magnitudes to a distance of $10$~Mpc.  For
the lower mass cases, the stars are brightest just when they collapse,
and only a narrow range of secondary masses can approach
the luminosity of the primary.  Most secondaries are fainter,
blue main sequence stars.
The secondaries of more massive stars are more visible for three
reasons.  First, for the same mass ratio they are simply more 
luminous.  Second, the 
luminosity differences between main sequence and evolved stars of
similar mass are smaller.  Finally, mass loss leads
to primary effective temperatures too high to efficiently radiate
in the optical.  For the \cite{Marigo2008} models
we see this only for the $M_p=40M_\odot$ case, but in the \cite{Eldridge2008}
models the $M_p=30M_\odot$ case would show the same properties.   

In Fig.~\ref{fig:frac1} we combine the distributions in secondary mass
and luminosity to examine the distribution of secondary magnitudes
relative to their primaries.  The $M_p=10$ and $20M_\odot$ 
primaries undergo core collapse as red stars close to the peak optical 
luminosity of all stars on the isochrone and generally have a blue companion.  
As a result, it is difficult for the companions to
significantly perturb the combined spectral energy distribution (SED)
that would be measured pre-explosion other than by creating a blue
excess.  For the $M_p=40M_\odot$ case, the secondaries are frequently
brighter than the primary for the reasons discussed above.  In general,
however, both stars are blue, so the overall SED will not be strongly
distorted by the combination.  While many of the secondaries are also
massive stars, most are relatively faint and can only be found directly
in relatively nearby galaxies ($<10$~Mpc).  

Table~\ref{tab:observe} summarizes several of these observational points
for $F=1$.  We first examined the contribution of the secondary to the blue flux
of the combined SED by computing the fraction of ccSN where a secondary
contributes more than 10\% or 100\% of the blue flux of the primary.
For lower mass stars, where the primary explodes as a red star, the
secondary generally distorts the SED and appears as a blue excess.
For the low mass and uniform secondary distributions, the secondary
is rarely (5--15\%) as bright in the blue as the primary, but
frequently (25--50\%) a significant ($>10\%$) perturbation.
Spectral distortions are far more common in the twins model.  Contamination
at redder wavelengths for these lower mass stars is minimal and will
have no effect on the inferences about Type~IIP progenitors by
\cite{Smartt2009}.
For the high mass $M_p=40M_\odot$ case, the secondary is frequently brighter
than the primary at all optical wavelengths because both stars are 
hot enough for their optical colors to be saturated.  The intermediate
$M_p=30M_\odot$ case is the least affected by the presence of a secondary
in the \cite{Marigo2008} models because the primary is both intrinsically
luminous and has retained a significant Hydrogen envelope.  Nonetheless,
many of the secondaries are intrinsically luminous and relatively 
easy to detect after the explosion fades.

We can also use these models to roughly estimate the fraction of
systems which will interact.  From the \cite{Marigo2008} models 
we can estimate the initial and maximum radius of the stars.
We distribute the semi-major axes $a$ uniformly in $\log a$ \citep{Kobulnicky2007}
from a minimum radius $a_{min}$ to $10^5 a_{min}$ where the 
minimum radius was set so that the initial stellar radii 
were $1/2$ the initial Roche lobe radius, following \cite{startrack}.  With these assumptions,
25--33\% of the primaries will fill their Roche lobes at their
maximum stellar radii.   The fraction depends  weakly on the dynamic range chosen 
for the semi-major axes,
rising to 30--45\% if we reduce it by an order of magnitude, and
dropping to 20-30\% if we raise it by an order of magnitude.  These
ranges are consistent with more detailed studies \citep[e.g.,][]{Podsiadlowski1992,
Kobulnicky2007,Eldridge2008}, and should
mean that adding interactions changes the details of our picture but
not the broad outline.

\section{Constraints}
\label{sec:observations}

Most accountings of the role of binaries in SN are based on the distribution
of supernova types \cite[e.g.,][]{Podsiadlowski1992,Kobulnicky2007,Eldridge2008}.  Suppose
stars from $M_{min} < M < M_{strip}$ explode as non-stripped Type~II SN
(IIP, IIL, IIn), and stars with $M>M_{strip}$ explode as stripped Type~IIb
and Ib/c SN where the fraction of stripped SN is about $f_{strip}\simeq 35\%$ and
$M_{min} \simeq 8M_\odot$ \citep{Smartt2009}.  In the absence of
any binary interactions and assuming a Salpeter mass function, producing
the stripped SN requires the stripping mass scale to be 
$M_{strip} = f_{strip}^{-0.74}M_{min} \simeq 2.2 M_{min}\simeq 18M_\odot$.  This
is much lower than the mass scale $M\simeq 25 M_\odot$ beyond which
models for massive stars lose their envelopes 
prior to core collapse \citep{Lejeune2001,Eldridge2008,Marigo2008}, although it is curiously similar to
the upper bound on Type~IIP progenitors found by \cite{Smartt2009}.
If we now allow fraction $f_i$ of the progenitors with $M<M_{strip}$ to 
interact and become stripped supernovae by mass transfer processes, then
the interacting fraction must be $f_i=0.08$, $0.17$ and $0.22$ for 
$M_{strip}=20$, $25$ and $30M_\odot$ in order to produce the observed
numbers of stripped SN, and  the fractions of stripped SN created by
binary interactions are 16\%, 38\% and 52\%.   While this is overly simplified, it is roughly
consistent with the detailed models of binary evolution, and, if only 
$\sim25\%$ of binaries are interacting, it requires the overall binary 
fraction $F$ to be high, consistent with \cite{Kobulnicky2007}.  
Note, however, that creating too large a fraction of stripped SN by binary
interactions of lower mass stars may by inconsistent with the
strong positional correlations between Type~Ib/c SN and H$\alpha$
emission from young massive stars \citep{Anderson2008}.  Stripped
ccSN due to binary evolution are likely to show the weaker
positional correlations of Type~II SN with H$\alpha$ emission.   

The failure to detect Type~Ib/c ``progenitors'' provides our first 
constraint.  Suppose our $M_p=40 M_\odot$ model is typical for
massive stars stripped by winds to become Type~Ib/c SN, so that the
progenitor detection limits are also strong limits on secondaries. 
Based on \cite{Smartt2009b}'s upper limits on 10 Type~Ib/c
progenitors of $M_R=-4.3$, $-5.7$, $-5.7$, $-6.0$, $-6.0$, $-6.1$, 
$-6.7$ $-6.9$, $-7.0$ and $-7.3$~mag, we can estimate the expected 
number of secondary detections for any primary mass and binary model.  
The constraint is relatively weak, but starting to constrain the
binary fraction from above.  For $M_p=40M_\odot$ and
$F=1$, the expected numbers of detectable secondaries in this sample
given the magnitude limits are $\sim 0.7$, $1.0$ and $0.5$ for the low 
mass, uniform and twins distributions respectively, corresponding to 30-50\%
probabilities of no detections.  In this case, the twins 
distribution has fewer detectable secondaries than the uniform
distribution because many of the twins are also optically faint
stripped stars (see Fig.~\ref{fig:mass}).  The results for any of the
other secondary distributions are similar -- only cases like the 
$F=1$, $M_p=30M_\odot$ twins distribution can be ruled out with 
high (95\%) confidence.  This stronger constraint for $30M_\odot$ 
arises from the lack of stripped secondaries at $30M_\odot$ compared
to $40M_\odot$ seen in Fig.~\ref{fig:mass}.  
However, a continued failure to
detect ``progenitors'' to Type~Ib/c, particularly with limits
as strong as those of \cite{Crockett2007} for SN~2002ap, will
begin to strongly constrain secondary distributions. 

The existence of some binaries does constrain the models from below,
although the limits will depend much on the definition of the problem.
Consider one example.  Assume there have been 22 nearby SN~II for 
which adequate pre-supernova imaging data exists -- the 20 SN from
\cite{Smartt2009} plus SN~1987A and SN~1993J.  In our present model,
where we have no binary interactions, both of these latter SN would
be normal Type~II SN.  Based on the binary models of SN~1993J, we 
compute the number of systems out of these 22 that would have a
secondary within 10\% (50\%) of the mass of the primary.  To have a 50\%
chance of at least one such system requires $F>0.73$, ($0.11$),  
$0.36$ ($0.07$) and $0.07$ ($0.05$) for the low mass, uniform and 
twins models, respectively.  Alternatively, if we interpret SN~1987A
as the SN of a secondary, then we must have $F>0.11$, $0.06$ and
$0.04$, respectively, to have a 50\% chance of having observed such
an event.  The constraints are again fairly weak, except that the
binary fraction must be very high for the low mass binary distribution to
have any likelihood of a system like SN~1993J.

Finally, we can consider the absence of V-band excesses at the level
of approximately 1~mag for the Type~IIP SN~2008bk and SN~2005cs, both
of which have estimated progenitor masses near $M_p\simeq 10M_\odot$. 
The fractions of ccSN with a secondary exceeding this limit are approximately
5, 9 and 31\% for the low mass, uniform and twins secondary distributions
and $F=1$.  Thus, while two such non-detections do not constrain
models, only a few additional, stronger or bluer limits are needed to begin
limiting the twins distribution.  

\section{Identifying and Verifying ccSN Secondaries}
\label{sec:verify}

There are two additional challenges for the problem of identifying
ccSN secondaries.  First, if searches are only feasible for $D<10$~Mpc,
the statistics will be poor and grow slowly unless the search can
be extended to historical ccSN where 
precise astrometric positions are lacking.  Second, once there
is a candidate secondary, we need some means of ruling out chance
alignments with an unassociated star.  We discuss the general 
problem of locating historical ccSN in the Appendix, while here
we focus on the solution to both of these problems for bright ccSN
secondaries.

Once the ccSN has faded, the developing SN remnant (SNR) enters
the ejecta-dominated phase.  Most of the ejecta is cool due to 
adiabatic expansion.  At the surface there is a thin layer of 
shocked CSM and SN material separated by a contact discontinuity
\citep[see, e.g.,][]{Chevalier1982,Chevalier1994}.  Thus, searches for young 
SNR are not very successful because they search for optical 
(emission line), radio or X-ray emission (see Appendix) created 
by this thin sheath, and this is generally not very luminous unless 
the CSM is very dense.  

Cool, largely neutral gas is best found by absorption, and this was dramatically
demonstrated by the discovery of the remnant of the Type~Ia SN~1885 in M31
by \cite{Fesen1989}.  SN~1885 is now easily visible 
as a $\sim 0\farcs8$ diameter absorption disk against the bulge of M31 at the 
wavelengths of high optical depth absorption lines like
 Ca~I (4600\AA), Ca~II H\&K, Fe~I (3021, 3720\AA), 
and Fe~II (2300-2600\AA) \citep{Fesen2007}.
The same effects will be seen for ccSN.  For example, if we take a representative 
Type~II model, the $12M_\odot$ S15A model from \cite{Woosley1995},
the Na, Ca and Fe yields are of order $9\times 10^{-4}$,
$1 \times 10^{-2}$ and $5 \times 10^{-2}M_\odot$, corresponding
surface number densities in the remnant of order
\begin{equation}
   \Sigma \sim \lbrace 2 , 20, 40\rbrace \times 10^{16}
       \left( { 5000\hbox{km/s} \over \hbox{v} } \right)^2
       \left( { 100\hbox{years} \over t } \right)^2
       \hbox{cm}^{-2}.
\end{equation}
for expansion rate $\hbox{v}$ and elapsed time $t$. In full, self-similar 
models of the eject-dominated phase, the expansion of the contact discontinuity
is slightly slower (radius $\propto t^{\sim 0.9}$) than this simple,
ballistic scaling \citep{Chevalier1982}.  If we integrate over the
SNR, then the material is spread over the full velocity range, so the
optical depth at line center is of order 
%\begin{equation}
%    \tau \sim \lbrace 50, 540, 1140 \rbrace \epsilon^{-1} x f  
%      \left( { \lambda \over 1000\AA }\right)
%      \left( { 5000\hbox{km/s} \over \hbox{v} } \right)^3
%      \left( { 100\hbox{years} \over t } \right)^2
%       = \tau_0 \left( { 5000\hbox{km/s} \over \hbox{v} } \right)^3
%      \left( { 100\hbox{years} \over t } \right)^2
%\end{equation}
\begin{equation}
       \tau = \tau_0 \left( { 5000\hbox{km/s} \over \hbox{v} } \right)^3
      \left( { 100\hbox{years} \over t } \right)^2
\end{equation}
where $\tau_0 \simeq 300x$, $1100x$ and $1800x$ for the Na~I  (5890, 5895\AA), 
Ca~I (4579-4585\AA) and Ca~II (3934, 3968\AA) lines, and $x$ is the
fraction in the appropriate ionization state.  The Fe~I complex near 3800\AA\ 
will have a similar optical depth.  Provided $x \gtorder 10^{-3}$ of the 
material is in the ground or first excited state, as is true of the
\cite{Chevalier1994} models, the SNR will be optically thick to these
absorption lines. 

The key question is the evolution of the ionization fraction over the
first millennium.  For the (probably) Type~Ia SN~1006, no Fe~I absorption
is observed, but there is modest Fe~II absorption corresponding to
a surface density of order $10^{15}$~cm$^{-2}$ \citep{Fesen1988}. Models
by \cite{Hamilton1988} of the SN~1006 remnant found that 
Fe~I is largely photoionized after a few 100 years 
while there is still a significant fraction
of Fe~II.  The Fe~I absorption in the Tycho SNR (SN~1572) is also
weak \citep{Ihara2007}, so it seems likely that the low ionization
states only survive for a few hundred years in Type~Ia SNRs.  There
is no comparable information for ccSN remnants.  The Crab remnant
(SN~1054) shows no strong, broad absorption lines \citep{Sollerman2000}
and we could find no results for Kepler's SN~1604 or Cassiopeia A
($\sim$~300 years).  

For the same physical size, ccSN remnants are likely to be in lower ionization 
states than Type~Ia remnants because their higher masses and smaller 
expansion rates mean they have more material to photoionize and higher 
recombination rates. However, we must also consider
the ionizing effects of the secondary (or any other central source). 
While a full photo-ionization calculation including estimates of line
emission is well beyond the scope of this paper, we did examine the
photoionization of a homologously expanding, homogeneous envelope of
pure H at the level of \cite{Osterbrock1974}.  We computed the time
scale at which an ejecta mass of $M_e$ expanding at $5000$~km/s at
the outer edge of the SNR is reionized given an effective ionizing 
flux from the secondary of $Q$~photons/s, assuming case~B recombination and a temperature
of $10^4$~K for the reionized material.  For $M_e=10M_\odot$ and
$\log Q = 49$ (an 06 star roughly), the reionization front reached 
the surface of the SNR after 150 years.  The reionization time
then scales linearly with changes in $M_e$ and $Q$.  Thus, even in
the presence of a very hot, luminous secondary, there will be an
extended period during which the bulk of the SNR is neutral, although
a hot secondary will typically reionize the SNR before the reverse
shock does so.  For hotter stars, the reionization
of He will be significantly more important than in a normal HII
region because of its relatively higher abundance, particularly
if the progenitor star was stripped prior to the explosion.  

The discovery of the SN~1885 SNR in absorption against the bulge
of M31 is difficult to apply to more distant SNR.  First, the 
remnant must be resolved in order to 
produce an observable signal.  The shell radius is approximately
\begin{equation}
   r_s = { \hbox{v} t \over D } = 0\farcs035 
       \left( { \hbox{v} \over 5000\hbox{km/s} } \right)
       \left( { t \over 100\hbox{years}} \right)
       \left( { 3\hbox{Mpc} \over D } \right)
\end{equation}
where $D$ is the distance.  When the remnant is unresolved,
the fractional drop in flux relative to an adjacent, unobscured wavelength is
\begin{equation}
   {\Delta F \over F } \simeq  0.5 f_{back}
       \left( { \hbox{v} \over 5000\hbox{km/s} } \right)^2
       \left( { t \over 100\hbox{years}} \right)^2
       \left( { 3\hbox{Mpc} \over D } \right)^2,
    \label{eqn:floss}
\end{equation}
where $f_{back}$ is the fraction of the local emission behind
the SNR and we have assumed an HST point spread function (PSF) radius of
$r_{psf}=0\farcs05 > r_s$.  The signal is heavily diluted by
the background light encompassed by the PSF but not behind
the SNR.  Once the remnant is old enough, 
$t \gtorder 100 (D/3\hbox{Mpc})(\hbox{v}/5000\hbox{km/s})$~years,
to be resolved ($2r_s > FWHM =0\farcs1 $), 
the fractional drop of $\Delta F/F = f_{back}$ is simply the fraction 
of background light for an optically thick transition. Second,
the change in surface brightness between a broad off-band (500\AA) 
and narrow on-band (50\AA) observation must be detectable. For
filters of equal efficiency, observations using the optimal 1:3
exposure time ratio for a 10:1 bandpass ratio and assuming perfect 
image subtraction (off-band$-$on-band) to produce a difference 
image of the absorbed flux, the noise in the difference image
is four times worse than that of an image through the broad 
filter with the same total integration time. Thus, the signal-to-noise
ratio for a resolved ($r_s>r_{psf}$) absorption disk is
\begin{equation}
   \left( { S \over N } \right)_{disk}
     \simeq  { f_{back} \over 4.2} \left( { S \over N } \right)_{cont}
            \left( { r_s \over r_{psf} }\right)
     \simeq f_{back} \left( { r_s \over r_{psf} }\right)
          \left( { t_{exp} \over 2500\hbox{s}}\right)^{1/2}
            10^{-0.4(\mu_B-22.5)}
\end{equation}
scaled by the continuum signal-to-noise ratio $(S/N)_{cont}$ for
the HST WFC3/F438W filter and an exposure time of approximately
one orbit.  Significant
detections against the unresolved emission in the disk of a galaxy
are challenging -- one needs old ($r_s$ several times $r_{psf}$),
foreground ($f_{back}\simeq 1$) systems in high surface brightness
($\mu_B \ltorder 22$~mag/arcsec$^2$) regions of the disk.  

Disks, however, are very different from bulges, because mean surface 
brightness has little meaning for nearby galaxies on angular scales 
where a significant fraction of the average surface brightness is 
due to individually detected stars.  For a resolved star behind 
the remnant, the signal-to-noise ratio is
\begin{equation}
   \left( { S \over N } \right)_{disk}
     \simeq  { 1 \over 4.2} \left( { S \over N } \right)_{cont}
     \simeq 10 \left( { t_{exp} \over 2500\hbox{s}}\right)^{1/2}
            10^{-0.4(B-24)}.
\end{equation}  
This is a far easier observation, and the probability of detection
is largely set by the probability of finding a star behind the remnant. 
This has enormous implications when searching for secondaries to ccSN
or trying to use secondaries to localize ccSN {\it because the
secondaries are always inside the remnant.}  Moreover, as pointed 
out by  \cite{Ozaki2006}, in the context of searches for binary
companions to the progenitor of the Type~Ia Tycho SN, their
absorption signature is unique.  Foreground stars show no
absorption, background stars show both red and blue-shifted
absorption and companions show only blue-shifted absorption.
This means that searches for bright ($\gtorder 26$~mag) companions
can be carried out in finite observing time ($\ltorder 10$~orbits)
for the hundreds of years between when the direct emission from the
SN/SNR ceases and when the remnant reionizes.

\section{Discussion}
\label{sec:discuss}

The current picture of massive stars and the creation of ccSN implies that
a large fraction ($\gtorder 50\%$) of SN progenitors are in stellar binaries
at the time of the explosion.  In many cases, the secondary stars are
detectable in either pre-explosion or late time observations.  In
pre-explosion images, the secondary to a Type~II SN is most observable
as a blue excess in the progenitor SED with 10\% excesses being relatively
common (25--50\%) and 100\% excesses being relatively rare (10\%-25\%).
Excesses become more common as the secondary mass distribution is skewed
to higher masses.   {\it The secondary to a Type~Ib/c
SN will frequently ($\sim 50\%$) be the more optically luminous star, 
because the stripped primary is radiating primarily in the ultraviolet.}
Direct observations of secondaries are, however, challenging, and are
most feasible for SN closer than $\sim 10$~Mpc.

There are, at present, no systematic attempts to survey for secondaries other
than \cite{Ryder2006}'s deliberate search for a secondary star to the Type~IIb
SN~2001ig.  The data that are available permit a binary fraction $F=1$, 
although the limits on Type~Ib/c progenitors \citep{Smartt2009b} and 
on color excesses for two Type~II progenitors \citep{Maund2005,Li2006,Mattila2008}
begin to weakly constrain $F\simeq 1$ for a ``twins'' distribution of 
secondaries where 50\% of secondaries have 90-100\% the mass of the primary.
As progenitor statistics improve due to the accumulation of both SN and
adequate archival data, the data will begin to constrain these models.  
It is worth, however, considering the types of observations needed to
constrain the secondary population.

In pre-explosion images, the challenge is to recognize that the SED is a
blend of two stars, as \cite{Aldering1994} found for SN~1993J.  For the
Type~Ib/c SN, this may be difficult because both stars will typically
be blue, but for Type~II SN the signature will generally be a blue 
excess to a red star.  To date, most detections of SN progenitors are
at best in two bands, allowing the determination of a luminosity and
either one intrinsic color or an extinction \citep[see][]{Smartt2009}.
Limiting the presence of a secondary will generally require at least
three bands.  Ideally, a complete three band Hubble Space Telescope (HST)
survey of the $\sim 40$ nearby ($\ltorder 10$~Mpc) galaxies that dominate
the local supernova rate ($\simeq 1$/year, see \citealt{Kochanek2008}) would provide an archival 
legacy for finding both progenitors and secondaries into the future
as well as for a broad range of other astrophysical studies.  We should
note, however, that the recent transients in NGC~300 and SN~2008S in
NGC~6946, whatever their underlying mechanism,  show that mid-IR observations 
are also crucial to understanding progenitor systems because in some
cases they are self-obscured by dust (see \cite{Prieto2008b} and related references). 

The most dramatic possibility in pre-explosion observations is the 
identification of an eclipsing binary as the progenitor system.
Very little is known about the time variability
of progenitor systems, with upper bounds $\sim 0.2-0.3$~mag for both SN~1987A
over its last century \citep[see][and references therein]{2004JAVSO..32...89P}
and for SN~1993J over a 150 day period in 1984 \citep{1995AJ....110..308C}.
Few local galaxies have been the subject of the relatively deep, long duration 
monitoring projects that would be needed to detect
eclipses in these (super)giant stars.  Discovery of even a single example
would have dramatic physical consequences, not only by making a direct
link between binary evolution and a resulting supernova, but because of
the enormous amount of extra physical information that can be derived from
the binary period and eclipse properties.  These will, of course, be rare,
but we note that for the parameters of the \cite{Maund2004}
model of SN~1993J, the probability of it having been an eclipsing binary
exceeds 10\%, with eclipse durations of roughly two years occuring every
10 years.  While challenging, the necessary data can be obtained by
ground based monitoring of nearby galaxies, as illustrated by the
eclipsing binary in Holmberg IX, a tidal dwarf companion to M81,
discovered by \cite{Prieto2008} using
the Large Binocular Telescope.  In addition to monitoring any SN 
progenitors, such data would also inventory all luminous eclipsing
binaries and variable stars (e.g., Cepheids) as well as providing the
basis for determining whether any massive stars die as failed
supernovae without a dramatic explosion \citep{Kochanek2008}.  A single
epoch, higher resolution HST survey is an important complement to any ground
based monitoring.  While difference imaging methods make it relatively 
easy to obtain light curves of variable sources from the ground even at 10~Mpc
(the AC signal),
stellar crowding means that HST is needed to provide absolute calibrations (the DC signal). 

{\it Once the SN has faded, it is relatively easy to find and confirm bright secondary 
stars of ccSN in galaxies closer than $D<10$~Mpc.}  
During its coasting phase, the SNR will have very high absorption 
optical depths in lines of Ca~I, Ca~II, Na~I, Fe~I, and Fe~II, as \cite{Fesen1989}
and \cite{Fesen2007} have already used to identify the SNR of the Type~Ia SN~1885 in M31,
unless something photoionizes the gas.  The method used for SN~1885, searching for a 
disk of absorption against the emission from unresolved background stars, is possible
only when the surface brightness is relatively high ($\gtorder 22$~mag/arcsec$^2$)
and after $\sim 10^2$~years when HST can resolve the SNR.  This approach may work
best for identifying historical ccSN associated with star clusters (e.g. like 
SN~2004am and SN~2004dj in the \cite{Smartt2009} sample).
On the other hand, any resolved star
brighter than $\ltorder 26$~mag that is either behind or inside the SNR 
should produce a detectable absorption signal.  The two cases can be distinguished
because background stars will show both red and blue-shifted absorption while secondaries
will show only blue-shifted absorption \citep{Ozaki2006}.  Moreover, if the 
search is carried out shortly after the SN fades, contamination by a background
star is very unlikely given the compactness (milliarcseconds after 10 years) of
the SNR. Fortunately, ccSN secondaries are generally blue stars at the wavelengths
likely to show strong absorption features, although some care will be needed for
stars with significant Balmer absorption in their atmospheres.  

Existing studies of reionization of SNR by either emission from shocked material near
the surface or ambient UV backgrounds \citep[e.g.][]{Chevalier1982,Hamilton1988,
Chevalier1994,Fesen2007} suggest that the bulk of an SNR should remain neutral for several
centuries but then reionize on time scales $\ltorder 10^3$ years, although the late-time
analyses of this problem have all focused on Type~Ia remnants.  These time
scales seem to be broadly consistent with the limited available data on absorption
in SNR.  The effect of secondary stars on the reionization of remnants seems never 
to have been considered.  Crude models suggest that hot, luminous O star secondaries
can photoionize an SNR from the inside on time scales of $10^2$-$10^3$~years.
This will create a very peculiar cross between an HII region 
with the secondary star as the photoionizing source, a planetary nebula, because of 
the very high He and metal abundances, and an SNR because of the high ejecta
velocities.  Full, dynamical, radiative transfer simulations will be needed
to characterize the observability of this phase.   The obvious trial scenario,
the late time spectra of SN~1993J \citep[e.g.][]{Maund2009}, is probably little
affected by this process because the secondary is a cooler B star.

The first step in a campaign to identify secondaries is to take all nearby ccSN
with accurate positions, particularly from post-explosion HST imaging, and 
reexamine the sites.  Ideally, one would start with systems for which late
time observations already exist (e.g. the 11 ccSN examined by \cite{Li2002},
although these are generally more distant than optimal) to ensure that 
any source is unlikely to be originating from the SN/SNR.
The next step is to use the absorption method to try to
confirm some of the candidates.  Here the obvious first cases to try are the 
candidate secondary to SN~2001ig identified by \cite{Ryder2006} and, once the 
SN has faded enough to examine absorption in the secondary without contamination 
from emission by the SNR, the secondary star of SN~1993J and (possibly) SN~2008ax.  
The numbers and properties of the secondary population will constrain the role of
binaries in the evolution of massive stars and the production of the various
classes of ccSN.  They also provide an independent
check of the models used to infer the properties of progenitors.
In a few cases where the secondary is sufficiently bright, it may be possible
to examine abundances in the ccSN ejecta.  In the early phases, even elements with
ejecta masses of $\sim 10^{-6}M_\odot$ (e.g., Barium) have significant 
absorption optical depths, although detection may be limited by 
Doppler smearing with stronger lines and inferences may be limited by ejecta clumping.
Identifying the secondary to a historical SN will also establish that the
SN was not a Type~Ia and set a lower bound to the mass of the primary.

\acknowledgements 

I would like to thank J. Beacom, R. Chevalier, A. Hamilton, J. Johnson, M. Pinsonneault, 
R. Pogge, J. Prieto, S. Smartt, K. Stanek, and T. Thompson for comments and discussion.
This research made use of the Sternberg Astronomical Institute supernova catalogs and
the NASA/IPAC Extragalactic Database (NED), which is operated by JPL/Caltech,
under contract with NASA. 

\appendix

\section{Other Approaches to Identifying Historical Supernovae}

In \S\ref{sec:verify} we introduced a simple means of identifying
bright secondaries to ccSN even if the ccSN location is poorly
known.  This method should work for several hundred years after
the SN unless something photoionizes the remnant.  The absorption
method will also work if the SNR lies in front of a sufficiently
bright background star or if it lies in a region of high surface 
brightness like a star cluster.  Here we discuss other possibilities
for identifying the positions of historical ccSN with high
precision (\cite{Smartt2009} argue for a goal of order
$10$~mas, albeit at higher typical distances than $10$~Mpc).
To provide a sense of the problem,  \cite{Barth1996} and 
\cite{Vandyk1999} examined the environments of past SN with HST imaging
data, but could only localize the SN with positional uncertainties
of 1-10~arcsec that made anything beyond environmental studies
impossible. 

The goal of most searches for SNR in external galaxies has been to
carry out more uniform surveys than are possible in the Galaxy.
Searches have used optical emission lines \citep[e.g.,][]{Matonick1997},
radio emission \citep[e.g.,][]{Lacey1997}, and X-ray emission 
\citep[e.g.,][]{Pannuti2007}.  The optical and radio surveys
do not find counterparts to historical supernovae unless they
had strong CSM interactions such as the examples we discuss below.
 Specific X-ray searches have been more
successful, with X-ray counterparts having been identified for
most post-1970 SN,\footnote{See http://lheawww.gsfc.nasa.gov/users/immler/supernovae\_list.html.}  
with the record being SN~1941C \citep{Soria2008}.  In general,
there is little overlap between the sources found by the three
approaches \citep[see, e.g.,][]{Pannuti2007}.  A further problem 
is that radio and X-ray detections may not lead to good enough 
{\it optical} astrometry in these crowded fields due to problems
in matching the reference frames with high accuracy.

Two additional possibilities are late time light curves and
dust echos.  The SN themselves fade relatively rapidly. For example,
SN~1987A was fainter than its progenitor by  $\sim 800$~days \citep{Suntzeff1991}.
\cite{Li2002} observed 11 ccSN (mostly Type~IIn) at late times. 
The four observed later than 1000 days had detections interpreted
to be the SN but spanning an interesting range for secondaries
($M_V \sim -4$ to $-13$~mag, see Fig.~\ref{fig:frac1}). Extrapolating
the light curves using the mean decline rates suggests that in most
cases the SN are fainter than $M_V \sim -4$~mag after 2000 days,
On these long time scales, the only internal source of energy
is $^{44}\hbox{Ti}$ which produces a maximum luminosity of
$L_{44} \simeq 10^3 (M_{44}/10^{-4} M_\odot)\exp(-t/86\hbox{years})L_\odot$
if all the decay energy is absorbed and 25\% of this if only
the positrons are absorbed \citep{Woosley1989}.  
Combined with a slope of only $0.013$~mag/year, this will
be challenging to detect decades later even if
the emission is concentrated in the optical/near-IR.
Some older ccSN are, however, trivial to find decades after the
event because of CSM interactions.  In particular, bright radio 
supernovae like SN~1979C \citep{Milisavljevic2009},
SN~1980K \citep{Fesen1999}, SN~1986J \citep{Milisavljevic2008} or 
SN~1993J \citep{Maund2004} are still being observed.  The
general distribution of supernovae from the nearly invisible
$^{44}\hbox{Ti}$ regime to the strong CSM interaction regime
is not known.   One result of searching for secondaries to all
ccSN with accurate post-explosion images, would be to clarify
these issues and provide strategies for the ccSN without such
data.

The last possibility we consider is searching for dust echos. 
Extragalactic dust echos have been imaged for the Type~IIpec SN~1987A \citep{Crotts1988}, 
the Type ~Ia SN~1991T \citep{Sparks1999}, the Type~IIb SN~1993J \citep{Sugerman2002,Liu2003},
the Type~Ia SN~1998bu \citep{Cappellaro2001}, the Type~Ia SN~1995E \citep{Quinn2006}
and the Type~Ia SN~2008X \citep{Wang2008}.  They were generally
found using difference imaging techniques, taking advantage
of the frequently superluminal pattern speeds of the echos.
While these detections are all relatively close in time to the SN,
\cite{Rest2005} have demonstrated that echos of 400-600 
year old supernovae can be seen in the LMC, some of which also have
superluminal apparent motions (up to $\hbox{v}_e=3c$).  The LMC echos
are resolved ($\Delta \simeq 2\farcs5$ across) and can have
relatively high surface brightnesses (up to 22~mag/arcsec$^2$).
There have also been ground based attempts to find dust echos
in more distant galaxies. \cite{Boffi1999} examined the
sites of historical SN for unusually blue emission due
to scattered light, finding 16 candidates among 64 nearby
SN, and \cite{Romaniello2005} searched for polarized 
emission near 4 SN in M~83 without success.

The problem with any search for dust echos is recognizing
them against the crowded stellar background.  The successful
detections have all taken advantage of the apparent motions
of echos and difference imaging multiple epochs to solve
this problem.  Here we note that
observations comparable to \cite{Rest2005} are possible for
older SN in nearby galaxies using HST, particularly if there
is dust in the foreground of the SN.  The outer angular scale 
of the echos is determined by the elapsed time and the 
foreground dust distribution, with
%\begin{equation}
%    r_{out} = 0\farcs2 \left( { t \over 10\hbox{years} \right)
%          \left( { 3\hbox{Mpc} \over D } \right) \left(
%            1 + { 2 z_{dust} \over c t } \right)^{1/2},
%\end{equation}
\begin{equation}
    r_{out} = 0\farcs54 \left( { t \over 10\hbox{years}} \right)^{1/2}
            \left| { z_{dust} \over 10\hbox{pc}} \right|^{1/2}
          \left( { 3\hbox{Mpc} \over D } \right) 
           \left| 1 + { c t \over 2 z_{dust}  } \right|^{1/2},
\end{equation}
where $z_{dust}$ is the distance of the dust along the line
of sight from the SN.  Thus, the angular scales for nearby
($D<10$~Mpc) somewhat old (decades to centuries) ccSN dust
echos are well-suited to HST.  Theoretical models of echos
have been explored by \cite{Sugerman2003} and \cite{Patat2005},
but we will focus on scaling from the observed properties of the 
many century old LMC echos.  The surface brightnesses of the
brighter LMC echos can be very high, but HST will somewhat
under resolve them in one dimension because their widths 
($\Delta \sim 0\farcs04 (3\hbox{Mpc}/D)$) will be somewhat 
narrower than HST's PSF.  Failure to resolve the echo in one 
dimension reduces its effective surface brightness as 
$2.5\log \Delta /FWHM$ where $FWHM\simeq 0\farcs1$ is the 
resolution of HST.  Even so, high surface brightness echos
at $D<10$~Mpc should be relatively easy to detect if the stellar 
backgrounds can be suppressed.  The existing HST studies have all used
HST pre-explosion data to subtract the stellar background
and render the echos easily visible.  This is not feasible for 
historical SN, but, the strategy used for the LMC 
detections will work.  The proper motion of an echo
with effective velocity $\hbox{v}_e$ is 
\begin{equation}
   \mu = 0\farcs021 \left( { \hbox{v}_e \over c } \right)
               \left( { 3\hbox{Mpc} \over D } \right)~\hbox{year}^{-1}
\end{equation} 
and it must move one resolution element, $\mu \Delta t = FWHM$
for difference imaging to be used.
The required time scale is relatively short
\begin{equation}
    \Delta t \simeq 5 \left( { c \over \hbox{v}_e } \right) 
             \left( { D \over 3\hbox{Mpc} } \right) \hbox{years}.
\end{equation} 
even in the absence superluminal apparent velocities.  This
can be tested for many SN as part of any late time search
for secondary stars or by revisiting the SN studied by
\cite{Barth1996}, \cite{Vandyk1999}, \cite{Li2002}.

\end{document}